\documentclass[aps,pra,amssymb,preprint, amsfonts,amsmath,showpacs, superscriptaddress]{revtex4-1}
 \usepackage{graphicx}
  \usepackage{amssymb}
  \usepackage{float}
 
\begin{document}
\title{A Relativistic Dynamical Collapse Model}
\author{Philip Pearle}
\affiliation{Emeritus, Department of Physics, Hamilton College, Clinton, NY  13323}
\begin{abstract}
 {A model is discussed where all operators are constructed from a quantum scalar field whose energy spectrum takes on all real values. 
The Schr\"odinger picture wave function depends upon space and 
time  coordinates for each particle, as well as an inexorably increasing evolution parameter $s$ which labels a foliation of space-like hypersurfaces.  The  model is constructed to be manifestly Lorentz invariant in the interaction picture. Free particle states and interactions are discussed in this framework. Then, the formalism of the CSL (Continuous Spontaneous Localization) theory of dynamical collapse 
  is applied.  The collapse-generating operator is chosen to to be the particle number space-time density. Unlike previous relativistically invariant models, the vacuum state is not excited.  The collapse dynamics depends upon two parameters, 
a parameter $\Lambda$  which represents the collapse rate/volume and a scale factor $\ell$.  A common example of collapse dynamics, involving a  clump of matter in a superposition of two locations, is analyzed. The collapse rate 
 is shown to be identical to that of non-relativistic CSL when the GRW-CSL choice of $\ell=a=10^{-5}$cm, is made, along with  $\Lambda=\lambda/a^{3}$  (GRW-CSL choice  
 $\lambda=10^{-16}s^{-1}$).  However, it is also shown that the change of mass  of a nucleon over the age of the universe is then unacceptably large.  The case where 
 $\ell$ is the size of the universe is then considered.  It is shown that the collapse behavior is satisfactory and the change of mass  over the age of the universe is acceptably small, when $\Lambda= \lambda/\ell a^{2}$.} 
 \end{abstract}
 \email{ppearle@hamilton.edu}
\maketitle

\section{Introduction}\label{sec1} 

	In the CSL (Continuous Spontaneous Localization) theory of dynamical collapse \cite{paper1}\cite{paper2}\cite{reviews}, an extra, anti-Hermitian  Hamiltonian is added to the usual Schr\"odinger equation.  This extra term depends upon a classical random field and upon a set of completely commuting ``collapse-generating" operators.  If the usual Hamiltonian $H$ is set equal to 0 so that  state vector dynamics is solely due to the extra term then, evolving under one or another random field, state vectors asymptotically approach one or another joint eigenstate of the 
collapse-generating operators. 

The random fields which accomplish this are high probability fields, where  
their probability is determined by the second equation of the theory, the Probability Rule. It  states that, at time $t$,  the probability of a particular random field  is $\sim\langle\psi,t |\psi,t\rangle$, where $|\psi,t\rangle$ is the state vector which evolved under that particular random field. 

The CSL theoretical structure described above can be applied to various problems.  For example, it has recently been applied to inflaton field fluctuation operators in the early universe\cite{JDP} so that a particular universe (presumably ours) is chosen by collapse dynamics, instead of the superposition of universes given by the standard theory.

But, its best known application is the non-relativistic CSL model, in which the collapse-generating operators are mass density operators averaged over a  mesoscopic distance 
$a$. This model also possesses a second parameter, the collapse rate $\lambda$. These parameters were first introduced by Ghirardi, Rimini and Weber in 	
the context of their SL (Spontaneous Localization) theory\cite{GRW}.

The dynamics  gives the Born Rule probabilities for the outcomes of a large class of experiments. Even when agreeing with the predictions of standard quantum theory in these cases, the theory nonetheless does something new:  its state vector 
describes the individual  outcome of an experiment (not the standard theory's superposition of such outcomes), allowing a correspondence between the state vector and a state of reality in nature.  However, there are experiments for which the theory makes different predictions than standard quantum theory.  So far, experiments along this line have not found a deviation from 
the predictions of standard quantum theory but neither have they found a deviation from the CSL theory.  They have put experimental limits on the two parameters 
of the theory\cite{Tumulka}. 

There is a long history of attempts to make a Lorentz invariant version of CSL\cite{attempts} using standard quantum field theory operators.  The problem has not been writing down such a model.  The problem has been that the extra term gives rise to particle production from the vacuum, 
\[
\frac{d}{dt}TrH\rho(t)\sim\int d{\bf x}\delta^{3}({\bf x}=0)=V\int d{\bf k}e^{i{\bf k}\cdot {\bf 0}}=V\int d{\bf k}, 
\]
\noindent ($Tr$ is the trace operation, $H$ is the Hamiltonian, $\rho(t)$ is the density matrix at time $t$, $\rho(0)=|0\rangle\langle0|$ is the density matrix at time 0, $V$ is the volume of space) because it excites each mode (corresponding to particles of momentum ${\bf k}$) in the vacuum equally. 

This amounts to infinite energy/sec-vol which, of course, is unacceptable.  One can see that a model must produce such an infinity if there are any particles at all produced out of the vacuum. For, if a particle of momentum ${\bf k}$ is produced in one reference frame, Lorentz invariance demands that another frame see the Lorentz transformed momentum ${\bf k}'$, but the equivalence of frames requires that momentum ${\bf k}'$ also appear in the first frame, ergo, all momenta are produced in any frame. Thus, a viable theory must have strictly 0 energy production from the vacuum. 	
	
Recently, Bedingham\cite{B} has shown how to construct a relativistic collapse model, where the energy production from the vacuum is zero. This model utilizes a non-standard quantum field $\phi(x)=\phi({\bf x}, t)$.  I first introduced this field\cite{PCCQ} to provide a quantum version of the  CSL c-number random field responsible for collapse.  The commutator between the 
field and its conjugate momentum is $i\delta^{4}(x-x')$, instead of the usual equal-time commutator $i\delta^{3}({\bf x}-{\bf x}')$.  Bedingham's model utilizes this field  as the collapse-generating operator, and also utilizes the usual CSL c-number random field.  This work is ingenious and valuable.  It not only provides a proof of concept, that a relativistic collapse model is not impossible, but it suggests that this field can be exploited to produce other relativistic models. 

In this paper, another relativistic model based upon $\phi(x)$ is presented.  Section II  discusses the properties of the field. Section III explores some aspects  of  a relativistically invariant  quantum theory based upon it.   Section IV adds  a collapse evolution of the state vector,  presents the resulting density matrix evolution equation, and shows that 
the vacuum state is unchanged by the evolution.  Section V treats a collapse example. Section VI treats a nucleon's mass change.

\section{Quantum Field}

The field $\phi(x)$ is constructed from annihilation and creation operators $a(k), a^{\dagger}(k)$ which correspond to particles of all possible real four-momenta:
\begin{equation}\label{1}
[a(k), a^{\dagger}(k')]=\delta^{4}(k-k').
\end{equation}
The one-particle state $|k\rangle\equiv a^{\dagger}(k)|0\rangle$ has  arbitrary four-momentum $({\bf k}, k^{0})$ but,  if one wishes, one can superpose such states to create  a normalized state 
whose four-momentum is narrowly spread around $({\bf k}, \sqrt{{\bf k}^{2}+m^{2}})$. 

From these operators, one can construct three fields of interest:
\begin{subequations}
\begin{eqnarray}\label{2}
\phi(x)&=&\frac{1}{\sqrt{2}(2\pi)^{2}}\int dk[e^{ ik\cdot x}a(k)+ e^{- ik\cdot x}a^{\dagger}(k)]       \label{A2a}                                                    \\
\dot{\phi}(x)&=&i\frac{1}{\sqrt{2}(2\pi)^{2}}\int dk[-k^{0}e^{ ik\cdot x}a(k)+ k^{0}e^{- ik\cdot x}a^{\dagger}(k)]    \label{A2b}  \\
\pi(x)&=&i\frac{1}{\sqrt{2}(2\pi)^{2}}\int dk[-e^{ ik\cdot x}a(k)+ e^{- ik\cdot x}a^{\dagger}(k)]
\end{eqnarray}
\end{subequations}
	\noindent where $dk=d{\bf k}dk^{0}$ (the unspecified integration range is $-\infty\leq k^{\mu}\leq \infty$), $k\cdot x={\bf k}{\bf {\cdot}}{\bf x}-k^{0}t$  and $\phi(x)\equiv \phi(x^{\mu})\equiv  \phi({\bf x},t)$.   Note that $\pi(x)\neq\dot{\phi}(x)$ as is the case in standard field theories. 
	
	The commutation relations are also rather unusual: 
\begin{eqnarray}\label{3}
[\phi(x), \phi(x')]&=&\int dk [e^{ ik\cdot (x-x')} - e^{- ik\cdot (x-x')}]=0   \hbox { so } [\phi(x), \dot {\phi}(x')]= [\dot\phi(x), \dot {\phi}(x')]=0\nonumber\\  
\ [\phi(x), \pi(x')]&=&i \delta^{4}(x-x').                                                       
\end{eqnarray}

 The Hamiltonian is
\begin{equation}\label{4}
H=\int dkk^{0}a^{\dagger}(k)a(k)=\frac{1}{2}\int dx[\dot\phi(x)\pi(x)+\pi(x)\dot\phi(x)]
\end{equation}
and one may readily verify that $[\phi(x),H]= i\dot\phi(x)$ and similarly for the other fields.  Likewise the total momentum operator is
\[
{\bf P}=\int dk{\bf k}a^{\dagger}(k)a(k)=-\frac{1}{2}\int dx[\nabla\phi(x)\pi(x)+\pi(x)\nabla\phi(x)].
\]
\noindent   The other generators of Lorentz transformations are $J^{r}\equiv i\epsilon^{rst}\int dk a^{\dagger}(k)k^{s}\frac{\partial}{\partial k^{t}}a(k)$ and 
$K^{r}\equiv i\int dk a^{\dagger}(k)[k^{0}\frac{\partial}{\partial k}+k^{r}\frac{\partial}{\partial k^{0}}]a(k)$.  One may then  show that $\phi(x)$ and its 
positive and negative frequency parts  transform like  Lorentz scalars, and $a(k), a^{\dagger}(k)$ transform like the Fourier transforms of a  
positive or negative frequency scalar field. It should be emphasized that, since all dynamical quantities shall be constructed from these field operators, the above generators are all that one needs to make Lorentz transformations, regardless of any interactions that might be introduced.

We define $\sqrt{2}\times$ the positive frequency part of $\phi(x)$: 
\begin{equation}\label{5}
\varphi(x)\equiv\frac{1}{(2\pi)^{2}}\int dke^{ ik\cdot x}a(k)\hbox{ which satisfies }[\varphi(x), \varphi^{\dagger}(x')]=\delta^{4}(x-x'), \medspace \varphi(x)|0\rangle=0.
\end{equation}
$\varphi^{\dagger}(x^{\mu})$, acting on the vacuum state, creates a particle at the event $x^{\mu}=({\bf x},t)$.

Now, consider the particle space-time number density operator $N(x)\equiv\varphi^{\dagger}(x)\varphi(x)$. Remarkably, its self-commutator vanishes everywhere:
\begin{eqnarray}\label{6}
[N(x),N(x')]&=&\varphi^{\dagger}(x)[\varphi(x),\varphi^{\dagger}(x')]\varphi(x')+\varphi^{\dagger}(x')[\varphi^{\dagger}(x),\varphi(x'])\varphi(x)\nonumber\\
&=&
\delta^{4}(x-x')[\varphi^{\dagger}(x)\varphi(x')-\varphi^{\dagger}(x')\varphi(x)]=0.
\end{eqnarray}
  \noindent $N(x)$ is a rather singular function, so it is useful to construct the less singular
  \begin{equation}\label{7}
\tilde{N}(x)\equiv\int dx'f[(x-x')^{2}]N(x').
\end{equation}
\noindent where $f$ is some useful function, with properties to be specified later: it is chosen to be dimensionless, so $\tilde{N}(x)$ is dimensionless since $N$ has dimension length$^{-4}$.   We note that, in addition to  $\phi$, and $\pi$, also $\varphi$, $N$ and $\tilde{N}$ all transform as Lorentz scalars.  The state of $n$ events is an eigenstate of $\tilde{N}(x)$:  
\begin{eqnarray}\label{8}
\tilde{N}(x)|x_{1}, x_{2} ... x_{n}\rangle &\equiv&\tilde{N}(x) \varphi^{\dagger}({\bf x}_{1}, t_{1})...\varphi^{\dagger}({\bf x}_{n}, t_{n}) |0\rangle=\nonumber  
\sum_{s=1}^{n}f[({\bf x}-{\bf x}_{s})^{2}-(t-t_{s})^{2}]|x_{1}, x_{2} ... x_{n}\rangle.\\
\end{eqnarray}
 
The quantum theory which naturally follows from the use of the operators defined above is different from the usual quantum theory. Here,  time is on an equal footing with space: wave functions are functions over all space-time.
Because $H$ has an infinite spectrum, one can define a mean time operator as the 0th component of a contravariant 4-vector:  
\begin{equation}\label{9}
\hat{X^{\mu}}\equiv \frac{\int dxx^{\mu} \varphi^{\dagger}(x)\varphi(x)}{\int dx\varphi^{\dagger}(x)\varphi(x)},  \hbox{ where }[\hat{X^{\mu}},P_{\nu}]=i\delta^{\mu}_{\nu}.
\end{equation}
A state consisting of n events is readily seen to be an eigenstate of the mean time operator:
\begin{equation}\label{10}
\hat{T}|{\bf x}_{1}, t_{1}; ..;  {\bf x}_{n}, t_{n}\rangle=\frac{1}{n}\int dxt \varphi^{\dagger}({\bf x}, t)\varphi({\bf x}, t)    \varphi^{\dagger}({\bf x}_{1}, t_{1})...\varphi^{\dagger}({\bf x}_{n}, t_{n})|0\rangle=\frac{t_{1}+ ... +t_{n}}{n}|{\bf x}_{1}, t_{1}; ..;  {\bf x}_{n}, t_{n}\rangle.
\end{equation}

 Just as $\exp-i{\bf P}\cdot{\bf a}$ provides a forward-in-space translation, so does $\exp i{\bf H}s$ provide a forward-in-time translation.  That is, since  
\begin{equation}\label{11}
e^{iHs}\varphi^{\dagger}({\bf x}, t)e^{-iHs}=\varphi^{\dagger}({\bf x}, t+s),
\end{equation}
\noindent it follows that
\begin{eqnarray}\label{12}
&&e^{iHs} |{\bf x}_{1}, t_{1}; ..;{\bf x}_{n}, t_{n}\rangle=e^{iHs}\varphi^{\dagger}({\bf x}_{1}, t_{1})...\varphi^{\dagger}({\bf x}_{n}, t_{n})|0\rangle \nonumber\\  
 &&\qquad\qquad=\varphi^{\dagger}({\bf x}_{1}, t_{1}+s)...\varphi^{\dagger}({\bf x}_{n}, t_{n}+s)|0\rangle= |{\bf x}_{1}, t_{1}+s; ..;{\bf x}_{n}, t_{n}+s\rangle                     
\end{eqnarray}
\noindent It is worth emphasizing that, usually, the energy operator generates time-translations in the Schr\"odinger Picture, but here it is the negative of the energy operator.

\section{A quantum theory based upon this field}

 It is useful  to first discuss a classical mechanics analog.  It is possible to construct a classical relativistically invariant  theory of interacting particles where, in addition to position and momentum of each particle, time and energy of each particle are also (conjugate) dynamical variables\cite{class}.
These variables are functions of a `universal evolution parameter' $s$.  The equations of motion are relativistically invariant Newton's Law-like equations, but the derivatives of the variables are with respect to $s$.   

We would like to regard the present model similarly.  We consider a given foliation of space-like hyper-surfaces $\sigma(s)$ labeled by an inexorably increasing evolution parameter $s$.   Just as the usual wave function has position coordinates for each particle, here the wave function has event coordinates for each particle:
\begin{equation}\label{13}
|\psi,s\rangle=\int dx_{1}...dx_{n}\chi(s;x_{1}...x_{n})\frac{1}{\sqrt{n!}}\varphi^{\dagger}( x_{1})...\varphi^{\dagger}( x_{n})|0\rangle
\end{equation}
\noindent where $dx_{i}\equiv d{\bf x}_{i}dt_{i}$, $\chi$ is a symmetric function of its arguments, and  $\int dx_{1}...dx_{n}|\chi|^{2}=1$ so $\langle\psi,s|\psi,s\rangle=1$.  

Under a 
Poincar\'e transformation,  the wave function transforms like a scalar, i.e., 
\begin{eqnarray}
|\psi,s\rangle'\equiv U|\psi,s\rangle&=&\int\prod_{i=1}^{n} dx_{i}\chi(s;..x_{i}..)\frac{1}{\sqrt{n!}}\prod_{i=1}^{n}\varphi^{\dagger}( \Lambda^{-1}(x_{i}+a))|0\rangle\nonumber\\
&=&\int\prod_{i=1}^{n} dx'_{i}\chi(s;..\Lambda x'_{i}-a..)\frac{1}{\sqrt{n!}}\prod_{i=1}^{n}\varphi^{\dagger}(x')|0\rangle
\equiv\int\prod_{i=1}^{n} dx'_{i}\chi'(s;.. x'_{i}..)\frac{1}{\sqrt{n!}}\prod_{i=1}^{n}\varphi^{\dagger}(x')|0\rangle,\nonumber
\end{eqnarray}
\noindent where $\chi'(s;.. x'_{i}..)=\chi(s;..x_{i}..)$.  The theory is relativistically invariant because, as shall be seen, the dynamical equations (in the interaction picture) are relativistically invariant and, although there is a `special' hypersurface foliation,  that foliation may be chosen arbitrarily.  In what follows, we shall not need general arbitrariness, so we shall suppose that the hypersurfaces are Lorentz hyperplanes in one `special' reference frame, and that the variable $s$ labeling the hyperplanes coincides with the coordinate $t$ in that frame.  The time variables of all particles in the initial state vector's wave function are to be peaked in the neighborhood of $\sigma(s_{0})$, in this case $t_{0}$.  We cannot know the value of $s$ but, if we did, if there was a clock which registered $s$,  the squared magnitude of the wave function at $s$ gives the probability density of the distribution of events of all particles.  

The un-normalized  event state $\varphi^{\dagger}({\bf x},t)|0\rangle=(2\pi)^{-2}\int dke^{-ik\cdot x}a^{\dagger}(k)|0\rangle$ contains all four-momenta.  
Since $k^{02}-{\bf k}^{2}$ can take on any real value, the event state may be considered as describing a superposition of particles of all masses, including imaginary masses, i.e. tachyons.  Negative $k^{0}$  may be considered as corresponding to an  antiparticle.   

   A  normalized state  of a single particle, localized near an event  $({\bf x},t)$  in this `special' frame (where expressions are simpler, but from which one may transform to other frames), 
\begin{eqnarray}\label{14}
|\psi\rangle&=&\int d{\bf x}'dt'\frac{1}{(2\pi\sigma^{2})^{3/4}} e^{-\frac{({\bf x}'-{\bf x})^{2}}{4\sigma^{2}}} \frac{1}{(2\pi\sigma'^{2})^{1/4}} e^{-\frac{(t'-t)^{2}}{4\sigma'^{2}}}                             
|{\bf x}',t'\rangle\nonumber\\
&=&
\int d{\bf k}dk^{0}e^{-ik\cdot x}\frac{1}{(\pi/2\sigma^{2})^{3/4}} e^{-{\bf k}^{2}\sigma^{2}} \frac{1}{(\pi/2\sigma'^{2})^{1/4}}e^{-k^{02}\sigma'^{2}} a^{\dagger}({\bf k},k^{0})|0\rangle, 
\end{eqnarray}
\noindent is also a multi-mass state. 

However, one can construct a state vector which is as close to an eigenstate of mass $m$ as desired,  for example,  
\begin{eqnarray}\label{15}
|\psi\rangle&=&
\int d{\bf k}dk^{0}e^{-ik\cdot x}\frac{1}{(\pi/2\sigma^{2})^{3/4}} e^{-[{\bf k}-{\bf p}]^{2}\sigma^{2}} \frac{1}{(\pi/2\sigma'^{2})^{1/4}}e^{-[k^{0}-\omega(\bf{k})]^{2}\sigma'^{2}} a^{\dagger}({\bf k},k^{0})|0\rangle\nonumber\\ 
&=&\frac{1}{(2\pi)^{3/4}}\int d{\bf x}'dt'K({\bf x}'-{\bf x}, t'-t)\frac{1}{(2\pi\sigma'^{2})^{1/4}}e^{-\frac{(t'-t)^{2}}{4\sigma'^{2}}}  |{\bf x}', t'\rangle,
\end{eqnarray}
 \noindent where $\omega(\bf{k})\equiv \sqrt{{\bf k}^{2}+m^{2}}$ and 
 \begin{equation}\label{16}
 K({\bf x}'-{\bf x}, t'-t)\equiv\int d{\bf k}\frac{1}{(\pi/2\sigma^{2})^{3/4}} e^{-[{\bf k}-{\bf p}]^{2}\sigma^{2}}e^{i{\bf k}\cdot({\bf x}'-{\bf x})-i\omega(\bf{k})(t'-t)}
 \end{equation}
\noindent is a positive energy  solution of the Klein Gordon equation, with mean momentum ${\bf p}$. 
 \noindent Although Eq.(\ref{16}) has in it the full time evolution of the wave packet,  translating and spreading, the 
 time-dependent gaussian in the full wave function (\ref{15}) keeps it in the neighborhood of $t'=t$, not translating or spreading.
 
 There is a tradeoff here: increasing the mass accuracy means increasing $\sigma'$, which in turn means that the wave function is  more spread out in time.  How much spread out in time?  The natural unit of time here is the time it takes light to cross  the reduced Compton length, $\lambdabar/c=\hbar/mc^{2}$: for a nucleon, where $\lambdabar\approx2\cdot10^{-14}$cm, 
 this is $\approx 7\cdot 10^{-25}$s.  Masses of nucleons are known to about 7 decimal places\cite{NIST}.  Therefore, if $\sigma'\gtrsim 10^{7}\hbar/mc^{2}\approx 7\cdot 10^{-18}$s, any mass spread would go  unnoticed.  Thus, an appropriately chosen initial state can have its particles quite  narrowly spread in time about the initial hyper-surface 
 and still have very well defined masses.

This state, as do all states, time-translates but does not evolve under the evolution operator $-H$: in the `special' frame this is
\begin{eqnarray}\label{17}
e^{ iHs}|\psi\rangle&=&\frac{1}{(2\pi)^{3/4}}\int d{\bf x}'dt'K({\bf x}'-{\bf x}, t'-t)\frac{1}{(2\pi\sigma'^{2})^{1/4}}e^{-\frac{(t'-t)^{2}}{4\sigma'^{2}}}  |{\bf x}', t'+s\rangle\nonumber\\
&=&\frac{1}{(2\pi)^{3/4}}\int d{\bf x}'dt'K({\bf x}'-{\bf x}, t'-t-s)\frac{1}{(2\pi\sigma'^{2})^{1/4}}e^{-\frac{(t'-t-s)^{2}}{4\sigma'^{2}}}  |{\bf x}', t'\rangle.
\end{eqnarray}
\noindent  That is, although the event coordinate's time is now centered on $t'=t+s$ instead of $t$, $K$'s time argument is still centered about 0.

H  contains no information about a particle's mass. To obtain free particle dynamics, one must add an evolution Hamiltonian which contains the necessary mass-information.  Defining $H_{m}\equiv\int dk\omega(\bf{k})a^{\dagger}(k)a(k)=\int dx\varphi^{\dagger}(x)
\sqrt{-\nabla^{2}+m^{2}}\varphi(x)=\int dx[\sqrt{-\nabla^{2}+m^{2}}\varphi^{\dagger}(x)]
\varphi(x)$ and the evolution operator ${\hat H}\equiv H_{m}-H$ (note: $[H_{m},H]=0$), the evolution is
\begin{eqnarray}\label{18}
|\psi,s\rangle&=&e^{-i{\hat H}s}|\psi\rangle=\frac{1}{(2\pi)^{3/4}}\int d{\bf x}'dt'K({\bf x}'-{\bf x}, t'-t)\frac{1}{(2\pi\sigma'^{2})^{1/4}}e^{-\frac{t'^{2}}{4\sigma'^{2}}} e^{-is\sqrt{-\nabla'^{2}+m^{2}}} |{\bf x}', t'+s\rangle\nonumber\\
&=&\frac{1}{(2\pi)^{3/4}}\int d{\bf x}'dt'K({\bf x}'-{\bf x}, t'-t+s)\frac{1}{(2\pi\sigma'^{2})^{1/4}}e^{-\frac{t'^{2}}{4\sigma'^{2}}}  |{\bf x}', t'+s\rangle\nonumber\\
&=&\frac{1}{(2\pi)^{3/4}}\int d{\bf x}'dt'K({\bf x}'-{\bf x}, t'-t)\frac{1}{(2\pi\sigma'^{2})^{1/4}}e^{-\frac{(t'-s)^{2}}{4\sigma'^{2}}}  |{\bf x}', t'\rangle.
\end{eqnarray}
\noindent Here,  the wave packet  translates and spreads as $s$ increases.

To introduce interactions between particles, one may first define a scalar field whose mass is close to $m$,  written variously as
\begin{eqnarray}\label{19}
\Phi_{m}(x)&\equiv&\int_{0}^{\infty}d\mu^{2}b_{\nu}(\mu^{2}-m^{2})\int dk\delta(k^{2}+\mu^{2})\Theta(k^{0})[a(k)e^{ik\cdot x}+ a^{\dagger}(k)e^{-ik\cdot x}]\nonumber\\    
&=&\int_{0}^{\infty}d\mu^{2}b_{\nu}(\mu^{2}-m^{2})\int \frac{d{\bf k}}{\omega_{\mu}}
[a({\bf k},\omega_{\mu})e^{i{\bf k}\cdot {\bf x}-i\omega_{\mu}t}+ a^{\dagger}({\bf k},\omega_{\mu})e^{-i{\bf k}\cdot {\bf x}+i\omega_{\mu}t}]\nonumber\\ 
&=&\int d{\bf k}\int_{|{\bf k}|}^{\infty} dk^{0}b_{\nu}(k^{02}-\omega_{m}^{2})[a(k)e^{ik\cdot x}+ a^{\dagger}(k)e^{-ik\cdot x}]
\end{eqnarray}
\noindent where $\omega_{\mu}\equiv\sqrt{{\bf k}^{2}+m^{2}}$ and $b_{\nu}(a)$ is a function concentrated about $a=0$, 
such that $\lim_{ \nu\rightarrow 0 }b_{\nu}^{2}(a)\rightarrow \delta (a)$ ($b_{\nu}$ is an approximation to a `square root of a delta function', for example, ($2\pi\nu^{2})^{-1/4} e^{-a^{2}/4\nu^{2}}$). 

The first line of ({\ref{19}) shows that $\Phi_{m}(x)$ is a Lorentz scalar. The second line shows 
three ways it differs from the usual three-field expression, i.e., it is mass-smeared,  it uses  $a({\bf k},\omega_{\mu})$ rather than the usual 
 $a({\bf k})$ and the integrand has  the scalar $d{\bf k}/ \omega_{\mu}$ in it rather than the usual $d{\bf k}/ \sqrt{2\omega_{\mu}}$.  The third line shows that 
 $\int_{|{\bf k}|}^{\infty} dk^{0}b_{\nu}(k^{02}-\omega_{m}^{2})a(k)$ is a kind of projection operator acting on $a(k)$.  In the limit $\nu\rightarrow 0$, this becomes the usual 
 $a({\bf k})/\sqrt{2\omega_{m}}$ since it yields the usual commutation relations:
\begin{eqnarray}\label{20}
 &&[\int_{|{\bf k}|}^{\infty} dk^{0}b_{\nu}(k^{02}-\omega_{m}^{2})a(k)f(k),\int_{|{\bf k}'|}^{\infty} dk'^{0}b_{\nu}(k'^{02}-\omega_{m}^{'2})a^{\dagger}(k')f'(k')]\nonumber\\
 &=&\delta({\bf k}-{\bf k}') \int_{|{\bf k}|}^{\infty} dk^{0}b_{\nu}^{2}(k^{02}-\omega_{m}^{2})f({\bf k}, k^{0}) f'({\bf k}, k^{0})\overrightarrow {_{\nu\rightarrow 0}} 
 \delta({\bf k}-{\bf k}') \frac{1}{2\omega_{m}}f({\bf k},\omega_{m} )f'({\bf k},\omega_{m} ),     
\end{eqnarray}
 \noindent where $f,f'$ are arbitrary functions. Thus, it appears that an interaction such as $g\Phi_{m}^{n}(x)$, i.e., an interaction picture evolution
\[
|\psi, s\rangle_{I}={\cal T}e^{-i\int_{s_{0}}^{s}dxg:\Phi_{m}^{n}(x):}|\psi, s_{0}\rangle,
\]  
 (${\cal T}$ is the time-ordering operator) will yield Feynman diagrams whose associated expressions 
 will be close in value to those of the usual $g\Phi^{n}(x)$ theory, for small $\nu$, since the diagram expressions are determined by these commutation relations.  We shall not pursue that any further here, since the aim of this paper is to explore collapse dynamics. 
 
 Before doing so however, one may mention another possibility for introducing interactions.  It is not readily possible to construct a Lorentz invariant interacting particle theory, 
 in analogy to the non-relativistic interacting particle theory, by adding a potential to the Klein-Gordon or Dirac equation Hamiltonians (which is one reason for introducing quantum fields).  
 However, it is possible within the peculiar framework we have discussed. One may introduce a Lorentz invariant two-particle interaction potential as follows.  If the Schr\"odinger picture state vector is related to the interaction picture state vector by 
$|\psi,s\rangle_{I}=e^{i\hat{H}s}|\psi,s\rangle_{S}$, let the evolution of the interaction picture state vector  be
\begin{equation}\label{21}
|\psi,s\rangle_{I}=e^{-\frac{i}{2}\int_{\sigma(s_{0})}^{\sigma(s)}dz\int dxdx'\varphi^{\dagger}(x)\varphi(x)V[z-x, z-x']\varphi^{\dagger}(x')\varphi(x')}|\psi,s_{0}\rangle.
\end{equation}
\noindent $V$ is a c-number potential which is a function of the three scalar products formed from the two four-vectors $z^{\mu}-x^{\mu}$ $z^{\mu}-x'^{\mu}$.  $x,x'$ are integrated over all space-time while $z$ is integrated between the two hypersurfaces. It follows from Eq.(\ref{6}) that there is no need for time-ordering.  We take $V(z-x,z-x)=0$ so that the exponent is equal to its normal ordered form.  
 
 The consequences of this possibility are peculiar.  The generic behavior is that particles of given approximate mass scatter into particles of a continuous distribution of mass, so this  
 does not correspond to our world. At best, with a 
suitably constrained $V$ and initial particle energies, one may arrange it so the outgoing particle masses are of the same approximate mass spread as the incoming particle masses. 
An example illustrating this behavior is discussed in Appendix A.

\section{Collapse Evolution}

We now turn to consider a model of collapse dynamics in the framework of such a quantum theory.  

The usual CSL collapse theory allows a realist interpretation (correspondence  of the state vector to the real physical world we see around us) because its dynamics causes  a `large' object in a superposition of places to become rapidly  localized to one place. By `localized' is meant that the center of mass part of the wave function is highly peaked at just one location, although it does extend over all space.  

Analogously, here, a realist interpretation is allowed because the dynamics causes a large object to become rapidly localized in space-time: the center of mass wave function is highly peaked at one event, although it does extend over all space-time.  

As is customary, we wish to demonstrate collapse behavior without the interference of additional dynamics. This is achieved by neglecting  $H_{m}$ and any interaction.  Thus, we set ${\hat H}=-H$. Then, without collapse, the particle states only translate in time in the Schr\"odinger picture.  We propose the following evolution equation in the interaction picture:

\begin{equation}\label{22}
|\psi, \sigma(s)\rangle=e^{-\frac{1}{4\Lambda}\int_{\sigma(s_{0})}^{\sigma(s)}dx{[w(x) -2\Lambda\tilde{N}(x)]^{2}}} |\psi, \sigma(s_{0})\rangle        
\end{equation}
\noindent (hereafter dropping the subscript I from the interaction picture state vector). In this equation, 
$\Lambda$ is a collapse rate parameter with dimension $L^{-4}$, $w(x)\equiv w({\bf x},t)$ is a c-number field of white noise structure (i.e., it can take any value at any event $x$) with the same dimension as $\Lambda$. 
 
That models of the form (\ref{22}) are Lorentz invariant was explained in references \cite{attempts}, but will be summarized here.  Under a Lorentz transformation, generated by applying a unitary transformation constructed from the ten Poincar\'e transformation generators given earlier, the effect on the exponential is to replace $\tilde{N}(x)$ by $\tilde{N}'(x)$, where the prime  denotes the quantity in the new reference frame. Relabeling the $x$ coordinates as $x'$ to obtain 
$\tilde{N}'(x')$ ($=\tilde{N}(x)$) changes $w(x)$ to $w(x')$. If this was instead  $w'(x')$ it would be manifestly Lorentz invariant, but it is not. However, the model considers the evolution of the set of state vectors under all possible $w$'s, and it is the set of all possible state vector evolutions which is Lorentz invariant. 

The dynamical equation (\ref{22}) has the well-known CSL form. It is supplemented by the Probability Rule which is that  the probability a particular random field 
 lies in the range $(w({\bf x},t),w({\bf x},t)+dw({\bf x},t) )$: 
 \begin{equation}\label{23}
P(w)Dw=Dw\langle\psi,\sigma(s)|\psi, \sigma(s)\rangle.   
\end{equation}
\noindent  where $|\psi, \sigma(s)\rangle$ has evolved from $|\psi, \sigma_{0}(s)\rangle$under that particular $w({\bf x},t) $. 

Some well known consequences of  Eqs.(\ref{22}),(\ref{23}) follow\cite{misc}.  Each individual state vector evolves toward one or another joint eigenstate of the 
competely commuting `collapse generating operators.' Here, these operators are $ \tilde{N}(x)$ for all $x$, and so the end product of collapse is a state of events $|x_{1}, x_{2} ... x_{n}\rangle$. Moreover, the probability distribution of the initial space-time density of events is preserved by the ensemble of evolutions.

It is immediately obvious that the vacuum state is unaffected by the evolution.   Set  $|\psi, \sigma(s_{0})\rangle= |0\rangle$ in (\ref{22}). Then, for  each state vector,  
\[
|\psi, \sigma(s)\rangle=C(s)e^{-\frac{1}{4\Lambda}\int_{\sigma(s_{0})}^{\sigma(s)}dxw^{2}(x)} |0\rangle\equiv C(s)|0\rangle,
\]
\noindent since 
$\tilde{N}(x)|0\rangle=0$. Thus, the problem of creation of particles out of the vacuum which afflicted previous attempts at achieving a relativistic collapse theory is, trivially, not a problem here.

We shall hereafter work in the `special' reference frame where  $\sigma(s)=s=t$.

Using Eqs.(\ref{22}) and (\ref{23}), one derives the density matrix evolution: 
\begin{equation}\label{24}
\rho(t)=  e^{-\frac{\Lambda}{2}\int_{t_{0}}^{t}dx[\tilde{N}_{L}(x)-\tilde{N}_{R}(x)]^{2}} \rho(t_{0}), 
\end{equation}
\noindent (the subscripts $L$ and $R$ mean that the operators are to be placed to the left or right of $\rho(t_{0})$) which satisfies the Lindblad equation 
\begin{equation}\label{25}
\frac{d}{dt}\rho(t)= -\frac{\Lambda}{2} \int d{\bf x}[\tilde{N}({\bf x},t),[\tilde{N}({\bf x},t), \rho(t)]].  
\end{equation}

These last two equations, which describe the behavior of the ensemble of state vectors, are all that shall be used hereafter. We note that, since the particle number space-time density 
operator $N(x)$ commutes with $\tilde{N}(x')$, it follows from (\ref{25}) that $dTr[\rho(t){\cal F}(N)]/dt=0$, where ${\cal F}$ is an arbitrary functional of  $N$.  Thus, the constancy of the probability distribution of the space-time density of events is verified.

\section{ Collapse}

A basic test for a collapse model is to apply it to an initial state $|\psi,0\rangle=\frac{1}{\sqrt{2}}[|L\rangle+|R\rangle]$ (initial density matrix $\rho(0)=\frac{1}{2}[|L\rangle\langle L|+|L\rangle\langle R|+|R\rangle\langle L|+|R\rangle\langle R|]$), where $|L\rangle$,  $|R\rangle$ 
describe an $n$ particle ``clump" of matter of volume $V$ whose centers are separated by a space-like interval $d^{2}\equiv |{\bf d}_{L}-{\bf d}_{R}|^{2}-(t_{L}-t_{R})^{2}>0$. 

To be concrete, we shall choose an expression for $f$ introduced in Eq.(\ref{7})'s definition of $\tilde{N}({\bf x},t)$.  It is  to have only space-like support:
\begin{equation}\label{26}
f(x^{2})=\frac{x^{2}}{\ell^{2}}e^{-\frac{x^{2}}{\ell^{2}}}\Theta(x^{2}),
\end{equation}
\noindent where $\Theta$ is the step function.  The reason we choose $f\sim x^{2}$ is to have it vanish smoothly on the light cone and thus be differentiable there. Possible choices for the scale $\ell$ shall be proposed shortly.  

Going to the reference frame where $t_{L}=t_{R}$, we shall consider that $|L\rangle$ consists of $n$ nucleons, with 
\begin{equation}\label{27}
|L\rangle=\prod_{j=1}^{n}\int dx_{j} \chi_{j}({\bf x}_{j}-{\bf d}_{L},t_{j})\varphi^{\dagger}({\bf x}_{j},t_{j})|0\rangle. 
\end{equation}
\noindent (For $|R\rangle$, replace 
${\bf d}_{L}$ by ${\bf d}_{R}$.)  For simplicity, we shall take the wave functions to have identical form, and be well localized with their `centers'  located at $({\bf x}_{j}={\bf z}_{j}+ {\bf d}_{L}, t_{j}=0)$. The ${\bf z}_{j}$ are sufficiently displaced from each other so that the wave functions  
\[
\chi_{j}({\bf x}_{j}-{\bf d}_{L,R},t_{j})=\chi'({\bf x}_{j}-{\bf d}_{L,R}-{\bf z}_{j})\frac{1}{(2\pi\sigma'^{2})^{1/4}}e^{-\frac{t_{j}^{2}}{4\sigma'{2}}}., 
\]
\noindent  are essentially non-overlapping.   The wave functions in the  left or right clump  are to occupy a volume $V$  with uniform density $D$. Thus, each clump  consists of  n uniformly distributed nucleii with, however, for simplicity, only one nucleon per nucleus. 

We now note: 
\begin{eqnarray}\label{28}
\tilde{N}(x)|L\rangle&=&\int dx_{1} f[(x-x_{1})^{2}]\varphi^{\dagger}(x_{1})\varphi(x_{1})|L\rangle\nonumber\\
&=& \int dx_{1} f[({\bf x}-{\bf x}_{1})^{2}-(t-t_{1})^{2}]\sum_{k=1}^{n}\chi_{k}({\bf x}_{1}-{\bf d}_{L},t_{1}) 
\varphi^{\dagger}(x_{1})|L\rangle_{k}\nonumber\\
&\approx&\sum_{k=1}^{n} f[({\bf x}-{\bf d}_{L}- {\bf z}_{k})^{2}-t^{2}] \int dx_{1}\chi_{k}({\bf x}_{1}-{\bf d}_{L},t_{1}) 
\varphi^{\dagger}(x_{1})|L\rangle_{k}\nonumber\\
&\approx&D\int_{V} d{\bf z}f[({\bf x}-{\bf d}_{L}- {\bf z})^{2}-t^{2}] |L\rangle.
\end{eqnarray}
\noindent where $|L\rangle_{k}$ is defined as the product (\ref{27}) without the $k$th term. The approximation in the third line considers that each nucleon's wave function's  support is very small in extent compared to $\ell$, so the argument $x_{1}$ in $f$ 
has been replaced by the `center' of   the wave function, $({\bf d}_{L}+{\bf z}_{k},0)$.  In the last line,  the sum over particles has been replaced by 
an integral over the particle number density. 

Eq.(\ref{28}) says that  $|L\rangle, |R\rangle$ are (approximate) eigenstates of $\tilde{N}(x)$.  Therefore, we may write Eq.(\ref{24}) as
\begin{eqnarray}\label{29}
\rho(T)&=&\frac{1}{\sqrt{2}}[|L\rangle\langle L|+|R\rangle\langle R|]+\frac{1}{\sqrt{2}}\big[|L\rangle\langle R|+|R\rangle\langle L|\big]e^{-I(T)}\hbox{ where}\nonumber\\
I(T)&\equiv&\frac{\Lambda}{2}D^{2} \int_{0}^{T}dx\Bigg[\int_{V} d{\bf z}
[f[({\bf x}-{\bf d}_{L}- {\bf z})^{2}-t^{2}]-f[({\bf x}-{\bf d}_{R}- {\bf z})^{2}-t^{2}]\Bigg]^{2}\nonumber\\
&=&\Lambda D^{2}\int_{0}^{T}dx\int_{V} d{\bf z}_{1}\int_{V} d{\bf z}_{2}f[({\bf x}- {\bf z}_{1})^{2}-t^{2}]\Big[ f[({\bf x}- {\bf z}_{2})^{2}-t^{2}] \nonumber\\
&&\qquad\qquad\qquad\qquad\qquad \qquad \qquad \qquad \qquad \qquad  - f[({\bf x}- {\bf z}_{2}-{\bf d})^{2}-t^{2}]\Big]\nonumber\\
&=&\Lambda D^{2}\int_{0}^{T}dx\int_{V} d{\bf z}_{1}\int_{V} d{\bf z}_{2}f[({\bf x}- {\bf z})^{2}-t^{2}]\Big[ f[{\bf x}^{2}-t^{2}] \nonumber\\
&&\qquad\qquad\qquad\qquad\qquad \qquad \qquad \qquad \qquad \qquad  - f[({\bf x}-{\bf d})^{2}-t^{2}]\Big],
\end{eqnarray}
\noindent where translation invariance of ${\bf x}$ has permitted writing the result in terms of ${\bf d}\equiv {\bf d}_{L}-{\bf d}_{R}$ and ${\bf z}\equiv {\bf z}_{1}-{\bf z}_{2}$. 

We shall  proceed to calculate (\ref{29}) for two cases; when $\ell=a\approx 10^{-5}$cm (the GRW-CSL scale parameter choice) and $\ell\approx$ ``size" of the universe, for the following reasons.

A new feature  compared to non-relativistic CSL is that a choice of distance scale ineluctably brings with it  a natural time $\ell/c$ which has dynamical consequences.     In the first case, $\ell/c$ is 
quite small, $a/c\approx 3\times 10^{-16}$s, in the second case quite large, $\ell/c\approx40\times10^{16}$s.  A laboratory collapse time
$T$  one  regards as satisfactorily small is still huge on the first scale, but is miniscule on the second scale. This requires two different estimates of (\ref{29}).  

Moreover, we have not yet discussed the increase of energy  and momentum of a particle due to the narrowing of the wave function by the dynamics. The mean value of the energy of any state does not change since, regarding  $\overline{H}(t)\equiv Tr H\rho(t)$, it is easy to see from Eq.(\ref{25}) that 
\begin{equation}\label{30}
\frac{d}{dt}\overline{H}(t)= -\frac{\Lambda}{2} \int d{\bf x}Tr\rho(t) [\tilde{N}({\bf x},t),[\tilde{N}({\bf x},t), H]] =-\frac{\Lambda}{2}\int d{\bf x}Tr\rho(t) [\tilde{N}({\bf x},t),i\dot{\tilde{N}}({\bf x},t)]=0.  
\end{equation}
\noindent In the first step, the trace operation allows the double commutator on $\rho(t)$ to be changed to a double commutator on $H$.  In the second step, we have 
used $[\tilde{N}({\bf x},t),H]=i\dot{\tilde{N}}({\bf x},t)$. The last step has used the vanishing commutator (\ref{6}). Similarly, the mean value of momentum, 
 $\overline{{\bf P}}(t)\equiv Tr{\bf P}\rho(t)$ does not change, since $[\tilde{N}({\bf x},t),\nabla{\tilde{N}}({\bf x},t)]=0$.

 Although the mean momentum of a particle does not change, the spatial narrowing of the wave function leads to an increase in the momentum spread: this also occurs in non-relativistic CSL.  There, the the momentum spread induces an energy spread since the energy operator is  the square of the momentum operator. However, there, the mass is a constant and so it is not affected by the dynamics.  Here,  there is not the relation between energy and momentum operators but the energy spread increases also because there is a 
concomitant narrowing in the time spread of the wave function.    

Thus,  a nucleon whose wave function is initially close to an eigenstate of mass $m$ gets smeared out in mass by the collapse evolution.  The smaller is $\ell$, the more rapidly do the energy spread and momentum spreads increase and the more rapidly does the mass change.  
To agree with experiment,  one would wish to have the mass $m$ of a nucleon change by no more than $\approx 10^{-7}m$ over the age of the universe.  As we shall see, this 
additional constraint beyond the requirement of a satisfactory behavior of $I$  makes the first choice, $\ell\approx a$, untenable, but permits the second choice.  

\subsection{Collapse when $\ell\approx a$}

For this case, $\ell\approx a$, we shall additionally assume that $a<<V^{1/3}<<d.$  

In this case we can argue that $f[({\bf x}-{\bf d})^{2}-t^{2}]$ in Eq.(\ref{29}) may be dropped, as the expression involving it makes a constant contribution for large $T$, while the expression involving the first term in the bracket 
gives a contribution $\sim T$.  This is most easily visualized in one spatial dimension, but holds in higher dimensions.  

Imagine the three forward light cones corresponding to the three arguments of $f$  in (\ref{29}), where the tips of cones 1, 2 and 3 are respectively at $t=0$ and $x=z, x=0, x=d$.  Imagine that the outside surface of each light cone has a coating of thickness $\approx a$: this is roughly the support of $f$ for each term in (\ref{29}).  

For large enough $T$, one can see that cones 1 and 3 intersect, so the product of the corresponding terms make a contribution to the integral, but that overlap does not change as $T$ increases, so neither does the contribution change.  

On the other hand, when $z=0$, cones 1 and 2 completely overlap and if $-a\lesssim z\lesssim a$ then they partially overlap.  Therefore, for $z$ in this range, there is an increase in the 
overlap integral $\sim T$ as $T$ increases. 

Thus, (\ref{29}) becomes

\begin{eqnarray}\label{31}
I(T)&\approx&\Lambda D^{2}\int_{0}^{T}dx\int_{V} d{\bf z}_{1}\int_{V} d{\bf z}_{2}f[{\bf x}^{2}-t^{2}]f[({\bf x}+{\bf z})^{2}-t^{2}].  
\end{eqnarray}
\noindent  Evaluation of (\ref{31})  is done in Appendix B, with the result 
\begin{eqnarray}\label{32}
I(T)&\approx&\frac{3\pi^{2}}{2}\Lambda Ta^{3}[DV][Da^{3}].
\end{eqnarray}
\noindent  Suppose we set 
\begin{equation}\label{33}
\Lambda a^{3}=\lambda,
\end{equation}
 where the latter is the GRW-CSL collapse rate parameter.  Since  the number of particles is $n=DV$ and, denoting 
the number of particles in a ``cell"  of volume $a^{3}$ by $n_{cell}$, we obtain the result
\begin{equation}\label{34}
I(T)\sim nn_{cell}\lambda T,
\end{equation}
\noindent which is identical to the collapse rate for the same situation in non-relativistic CSL. 

\subsection{Collapse for very large  $\ell$}

Here we make no constraint on the relative sizes of the clump and the clump separation $d$, but both are very small compared to $\ell$, and  $T<<\ell/c$.

We calculate (\ref{29})  in Appendix B, Eq.(\ref{B6}): 
\begin{equation}\label{35}
I(T)\sim n^{2}\Lambda\ell  T d^{2}.  
\end{equation}
\noindent If we choose 
\begin{equation}\label{36}
\Lambda\ell\approx \frac{\lambda}{a^{2}}, 
\end{equation}
\noindent then the collapse rate (\ref{35}) becomes
\begin{equation}\label{37}
I(T)\sim n^{2}\lambda T\Big[ \frac{d}{a}\Big]^{2}.  
\end{equation}
\noindent The non-relativistic CSL collapse rate for $n$ particles in a volume $\lesssim a^{3}$, separated by  distance $d\gtrsim a$, is $n^{2}\lambda T$.  
The result obtained here is comparable at $d\approx a$, but is faster  as $d$ increases.  It appears that the rate obtained here is as fast or faster than 
the non-relativistic rate for most situations of interest.   

\section {Change of  mass of a nucleon }

To characterize the mass spread of a nucleon, one might proceed in the following way.  First, define a hermitian mass operator $M\equiv\sqrt{|H^{2}-{\bf P}^{2}|}$.  Already there is a complication: since nothing forbids tachyonic 
four-momenta, there are negative eigenvalues of $H^{2}-{\bf P}^{2}$, which necessitates the absolute magnitude sign in the definition of $M$: otherwise, it would not be a Hermitian operator. Its standard deviation $\Delta$ is given by : 
 \begin{eqnarray}\label{38}
\Delta^{2}&\equiv&\overline{\langle\psi|(M-\langle{M}\rangle)^{2}|\psi\rangle}=
\overline{\langle\psi| |H^{2}-P^{2}||\psi\rangle}-\overline{\langle\psi|M|\psi\rangle^{2}},
\end{eqnarray}
\noindent where the overline refers to the ensemble average over all state vectors, each state vector $|\psi\rangle_{w}$ characterized by its evolution under a specific classical field $w(x)$.

A second complication is that the second term on the right hand side of \ref{38}), 
$\overline{\langle M\rangle\langle M\rangle}$ is not readily calculable because it is quartic in the state vector.  Thus, we consider the first term on the right hand side of ({\ref{38}), 
  which can be calculated using the density matrix, as an upper limit on $\Delta^{2}$.  However, the absolute magnitude complicates that calculation. But, if the mass starts out narrowly spread and positive and does not change much over the time interval of interest (a condition we shall impose), it is reasonable to suppose that, for the majority of state vectors of large measure, that 
 $_{w}\langle\psi| (H^{2}-P^{2})|\psi\rangle_{w}>0$ and so $\overline{\langle\psi| |H^{2}-P^{2}||\psi\rangle}\approx|\overline{\langle\psi| H^{2}-P^{2}|\psi\rangle}|$.
  Thus, we suppose 
 \begin{eqnarray}\label{39}
\Delta^{2}(T)\lesssim |\overline{H^{2}}(T)-\overline{P^{2}}(T)|.
\end{eqnarray}
 
 To calculate $\frac{d}{dt}\overline {H^{2}}$ we utilize ({\ref{25}):
 
\begin{equation}\label{40}
\frac{d}{dt}\overline {H^{2}}=-\frac{\Lambda}{2} Tr\rho(t)\int d{\bf x} [\tilde{N}({\bf x},t), [\tilde{N}({\bf x},t),H^{2}]]
= \Lambda Tr\rho(t)\int d{\bf x} \dot{\tilde{N}}({\bf x},t) \dot{\tilde{N}}({\bf x},t) .
\end{equation}

The right hand side of this equation is readily evaluated. Put  ({\ref{24})'s expansion for  $\rho(t)$ into ({\ref{40}).   Because of the trace, the operators 
$(1/4\Lambda)\int_{0}^{t}dx'[\tilde{N}(x')[\tilde{N}(x'), ]]$ in the exponent may be turned from operating on $\rho(0)=|\psi,0\rangle\langle\psi,0|$ to operating on 
 $\dot{\tilde{N}}({\bf x},t) \dot{\tilde{N}}({\bf x},t)$.  These commutators all vanish, so we are left with
\begin{equation}\label{41}
\frac{d}{dt}\overline {H^{2}}
= \Lambda \langle \psi,0|\int d{\bf x} \dot{\tilde{N}}({\bf x},t) \dot{\tilde{N}}({\bf x},t)|\psi,0\rangle .
\end{equation}
We wish to consider the initial wave function of just one nucleon (if we consider $n$ nucleons with non-overlapping wave functions, the result is just $n$ times larger), 
$|\psi,0\rangle=\int dx\chi(x)\varphi^{\dagger}(x)|0\rangle$.  Putting this into ({\ref{41}), and evaluating the matrix element gives
\begin{eqnarray}\label{42}
	\frac{d}{dt}\overline {H^{2}}&=& \Lambda \int d{\bf x} \int dx_{1} |\chi(x_{1})|^{2}\dot f ^{2}[(x-x_{1})^{2}] = \Lambda \int d{\bf x} 
	 \int dx_{1} |\chi(x_{1})|^{2}\dot f ^{2}[{\bf x}^{2}-(t-t_{1})^{2}]\nonumber\\
	&\approx&\Lambda \int d{\bf x} \dot f ^{2}[{\bf x}^{2}-t^{2}].
\end{eqnarray}
\noindent Here we have first used translation invariance in ${\bf x}$ to remove ${\bf x}_{1}$ from the argument of $f$.  Then,  as  previously, we have assumed that the wave function's center of time coordinate is $x_{1}^{0}\equiv t_{1}=0$ and  that the wave function is well localized in time on the scale of $\ell$ so that $f$ scarcely changes over the time spread of  
$|\chi(x_{1})|^{2}$.  Finally, we have utilized $\int dx_{1}|\chi(x_{1})|^{2}=1$.

A similar calculation, for $\overline{P^{2}}$ gives,
\begin{equation}\label{43}
	 \frac{d}{dt}\overline {P^{2}}= \Lambda \int d{\bf x} (\nabla f) ^{2}[{\bf x}^{2}-t^{2}].
\end{equation}

Define $f'(s^{2})\equiv d/ds^{2}f(s^{2})=\ell^{-4}(\ell^{2}-s^{2})\exp-\ell^{-2}s^{2}$, and then it follows from ({\ref{39}),  ({\ref{42}) and ({\ref{43}) that
\begin{eqnarray}\label{44}
\Delta^{2}(T)&\lesssim&4\Lambda\mu^{4}\int_{0}^{T}dt \int d{\bf x} f'^{2}[{\bf x}^{2}-t^{2}]|t^{2}-{\bf x}^{2}|+\Delta^{2}(0)\nonumber\\
&\lesssim&\frac{16\pi\Lambda}{\ell^{8}}\int_{0}^{T}dt\int_{t}^{\infty}r^{2}dr[\ell^{2}-(r^{2}-t^{2})]^{2}(r^{2}-t^{2})e^{-2\ell^{2}(r^{2}-t^{2})} +\Delta^{2}(0).
\end{eqnarray}
\noindent Now, change to hyperbolic variables $r=s\cosh\beta$, $t=s\sinh\beta$, and $dtdr=sdsd\beta$:
\begin{eqnarray}\label{45}
\Delta^{2}(T)-\Delta^{2}(0)&\lesssim&\frac{16\pi\Lambda}{\ell^{8}}\int_{0}^{\infty}s^{5}ds[\ell^{2}-s^{2}]^{2} e^{-2\ell^{-2}s^{2}} \int_{0}^{\sinh^{-1}T/s}  d\beta \cosh^{2}\beta.               
\end{eqnarray}

\subsection{ $\ell\approx a$}

Since $T/s\simeq T/a>>1$, the upper limit on the $\beta$ integral is $\approx \ln 2T/s$. The resulting integrals are then elementary (or, one obtains the same result by making the approximation 
$\cosh\beta\approx 2^{-1}e^{\beta}$) with the result:
\begin{equation}\label{46}
\Delta^{2}(T)-\Delta^{2}(0)\lesssim2\pi\Lambda T^{2}.
\end{equation}

Let the initial spread of mass be $\Delta(0)\approx10^{-7}m$, so the nucleon mass is known to present-day accuracy.  We ask how large $T$ is for the
change of $\Delta$ to be of similar accuracy, $\Delta(T)-\Delta(0)\approx10^{-7}m$.  With $\Lambda\approx\lambda/a^{3}$ (the choice that made the collapse rate essentially 
the same as for non-relativistic CSL), we have, from ({\ref{46}), using $a^{-1}\equiv\mu\approx 2$eV:
\begin{eqnarray}\label{47}
\frac{\Delta(T)-\Delta(0)}{m}&\lesssim&\frac{\pi\Lambda T^{2}}{\Delta(0)}m
 =\pi(\lambda T)\frac{T}{a}\frac{\mu^{2}}{\Delta(0)m}\nonumber\\
 &\approx&\pi(10^{-16}T)\frac{3\times 10^{10}T}{10^{-5}}\frac{1}{10^{-7}}\Big[\frac{2}{940\times 10^{6}}\Big]^{2}
 \approx10^{-7}\Big[\frac{T}{50}\Big]^{2}.
\end{eqnarray}
\noindent where $T$ is in seconds.  Thus, in about a minute the limit is reached whereas, to agree with experiment, the limit should at best be reached only over the age of the universe.  

As might be expected, the energy production rate is unacceptably large as well.  
            
\subsection{Very large  $\ell$}

Returning to Eq.({\ref{45}), set $\ell ={\cal T}$, the age of the universe,.   Since $T/s\simeq T/{\cal T}\lesssim 1$,  the upper limit on the $\beta$ integral  may be approximated as $\sinh^{-1}(T/s)\approx T/s$.
With $\cosh\beta\approx 1$, after performing the integral over $s$,  the result is:
\begin{equation}\label{48}
\Delta^{2}(T)-\Delta^{2}(0)\lesssim.2\Lambda\ell T.
\end{equation}
\noindent With the choice ({\ref{36}), $\Lambda\ell\approx \lambda/a^{2}$ (curiously, in both the calculation of collapse rate and of mass spread increase it is a single parameter, 
$\Lambda\ell$, which determines the results) we have for $T={\cal T}$:
\begin{equation}\label{49}
\frac{\Delta(T)-\Delta(0)}{m}\lesssim .1(\lambda {\cal T})\frac{\mu^{2}}{\Delta(0)m}\approx2\times10^{-12}.
\end{equation}

We conclude that this case gives satisfactory results for the mass-spread increase.

The energy production rate is likewise satisfactory.  From Eq.({\ref{42}) it follows that 
\begin{eqnarray}\label{50}
\overline {H^{2}}(T)-\overline {H^{2}}(0)&=& \Lambda \int_{0}^{T}dx \dot f ^{2}[{\bf x}^{2}-t^{2}]\nonumber\\
&=&\frac{16\pi\Lambda}{\ell^{8}}\int_{0}^{T}dtt^{2}\int_{t}^{\infty}r^{2}dr[\ell^{2}-(r^{2}-t^{2})]^{2}e^{-2\ell^{-2}(r^{2}-t^{2})}\approx 6\Lambda\ell^{-1}T^{3}. 
\end{eqnarray}
\noindent Taking $\overline{H^{2}}(0)\approx m^{2}$, then
\begin{equation}\label{51}
\sqrt{\overline {H^{2}}(T)-m^{2}}\approx 3\mu\lambda T \Big[\frac{T}{{\cal T}}\Big]^{2}\frac{\mu}{m}\approx 10^{-6}\hbox {eV}
\end{equation}
\noindent for $T={\cal T}$.

To conclude, the model presented here appears to provide a satisfactory  relativistic generalization of the non-relativistic CSL collapse model, although there remains much  
to be explored.

\acknowledgments{I am very grateful to Daniel Bedingham,  Adam Bednorz and Daniel Sudarsky for their comments. }

\appendix

\section{Two particle scattering example}

We consider here the evolution, given in  Eq.({\ref{21}). We specialize the scalar potential to be 
\begin{equation}\label{A1}
V[z-x, z-x']=W[(x-x')^{2}]\Theta[(z-x)^{2}]\Theta[(z-x')^{2}]\Theta[-(2z-x-x')^{2}]\equiv W[(x-x')^{2}]{\cal I}(z;x;x')
\end{equation}
\noindent where $W$ has finite range, i.e., it vanishes beyond a specified value of its argument. .  The reason for the step functions limiting  $z-x,z-x'$ to space-like values and $z-(x+x')/2$ to time-like values, shall appear presently.

Applied to a two-particle state vector (the results are easily generalized to $n$ particles) in the `special' frame, with $s_{0}=0$ for simplicity, we have
\begin{eqnarray}\label{A2}
|\psi,s\rangle_{I}&=&e^{-\frac{i}{2}\int_{0}^{s}dz\int dxdx'\varphi^{\dagger}(x)\varphi(x)V[z-x, z-x']\varphi^{\dagger}(x')\varphi(x')} \int dx_{1}dx_{2}\chi_{0}(x_{1}, x_{2})\frac{1}{\sqrt{2}}\varphi^{\dagger}(x_{1})\varphi^{\dagger}(x_{2})|0\rangle           \nonumber\\
&=&\int dx_{1}dx_{2}e^{-iW[(x_{1}-x_{2})^{2}]\int_{0}^{s}dt\int d{\bf z}{\cal I}(z;x_{1};x_{2})}
\chi_{0}(x_{1}, x_{2})\frac{1}{\sqrt{2}}\varphi^{\dagger}(x_{1})\varphi^{\dagger}(x_{2})|0\rangle\nonumber\\
&&\negmedspace\negmedspace\negmedspace\negmedspace\negmedspace\negmedspace\negmedspace\negmedspace
\negmedspace\negmedspace\negmedspace\negmedspace\negmedspace\negmedspace\negmedspace\negmedspace
\approx\int dx_{1}dx_{2}e^{-iW[({\bf x}_{1}-{\bf x}_{2})^{2}]\int_{0}^{s}dt\int d{\bf z}{\cal I}({\bf z},t;{\bf x}_{1},0;{\bf x}_{2},0)}
\chi_{0}(x_{1}, x_{2})\frac{1}{\sqrt{2}}\varphi^{\dagger}(x_{1})\varphi^{\dagger}(x_{2})|0\rangle
\end{eqnarray}
\noindent where, in the approximation, we are assuming $\chi_{0}(x_{1}, x_{2})\sim e^{-[t_{1}^{2}+t_{2}^{2}]/\sigma'^{2}}$, with $\sigma'$ small enough that we may set $t_{1}\approx 0, t_{2}\approx 0$.

We see that the result of the interaction is that the wave function is just multiplied by a phase factor,  so the spatial distribution of events $|\chi_{0}(x_{1}, x_{2})|^{2}$ is unaffected by the interaction.  However, the four-momentum distribution \textit{is} affected, provided the initial wave packets described by $\chi_{0}$ are not so far apart that $W[({\bf x}_{1}-{\bf x}_{2})^{2}]$ vanishes on their support. 

Because of the step functions, the interaction has a finite range in time as well.  In one space dimension, the region of intersection of the three step functions is shown in Fig. 1.
\begin{figure}[h]
\includegraphics[width=3.5in]{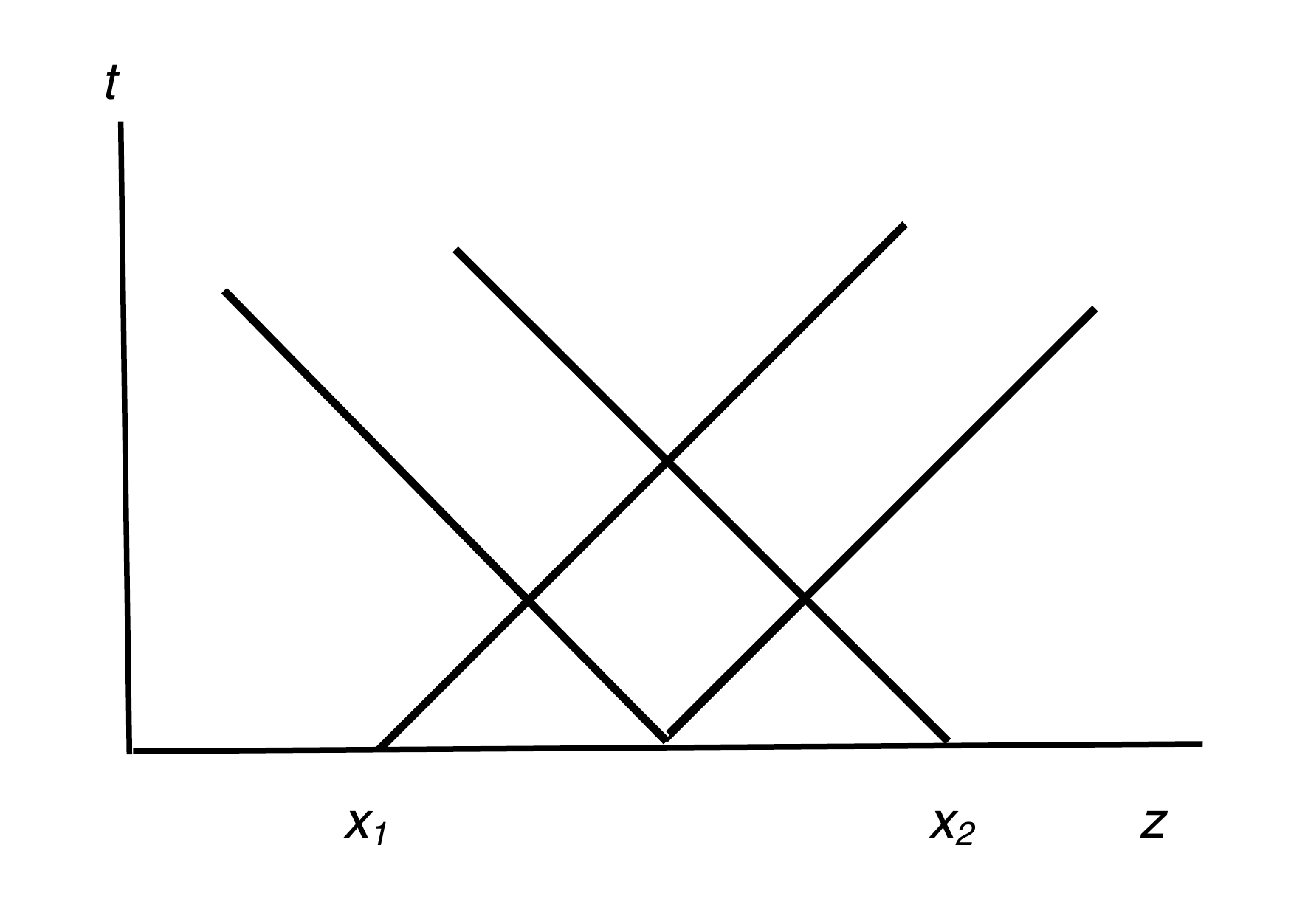}
\caption{ The  region where ${\cal I}({\bf z},t;{\bf x}_{1},0;{\bf x}_{2},0)=1$ is the  diamond shaped region, with vertices 
$(z,t)= (\frac{x_{1}+x_{2}}{2} ,0),( \frac{x_{1}+x_{2}}{2} +\frac{x_{2}-x_{1}}{4},\frac{x_{2}-x_{1}}{4}),(\frac{x_{1}+x_{2}}{2}, \frac{x_{2}-x_{1}}{2}), 
( \frac{x_{1}+x_{2}}{2} -\frac{x_{2}-x_{1}}{4},\frac{x_{2}-x_{1}}{4})$.}
\label{Fig.1}
\end{figure}  
Thus, $s$ increases to a value beyond which the interaction is ``turned off." 
 That is, the phase factor no longer changes (the potential vanishes) when $s$ exceeds a certain value.  This value is the maximum value of $t$ at the top of the diamond-shaped region 
when $|{\bf x}_{1}-{\bf x}_{2}|$ equals the range of $W$.  If the potential was not turned off, e.g., if the exponent was instead $\sim sW$, this would effectively be a potential whose coupling constant grows without bound, which in turn would give rise to energy-momentum changes without bound in the wave-function. 

Such a turning off of potential is a property of non-relativistic (Galilean invariant) quantum theory.  There, in the Schr\"odinger picture, a potential operator $W[({\bf X}_{1}-{\bf X}_{2})^{2}]$ doesn't explicitly change with time, but its effect  is eventually turned off because the wave packets, which initially move freely, eventually pass through the interaction region, and end up  moving freely beyond the range of the potential.  The same behavior of course must be seen in the interaction picture.  There, while the packets do not move 
freely, the potential $W[({\bf X}_{1}+{\bf P}_{1}t/m-{\bf X}_{2}-{\bf P}_{2}t/m)^{2}]$  now has dynamical dependence. The argument of the potential effectively eventually grows larger than the potential range, which turns off the potential.  

In the model discussed here, there is a reversal of what happens in these pictures: non-dynamical behavior of the potential belongs to the interaction picture potential (the form exhibited in  ({\ref{20}), necessitated by the constraint of relativistic invariance),  while it is in the Schr\"odinger picture that the potential has dynamical dependence:
\begin{eqnarray}\label{A3}
&&|\psi,s\rangle_{S}= e^{- i{\hat H}s}|\psi,s\rangle_{I}\nonumber\\
&&=
\int dx_{1}dx_{2}e^{-iW[({\bf x}_{1}-{\bf x}_{2})^{2}]\int_{0}^{s}dt\int d{\bf z}{\cal I}({\bf z},t;{\bf x}_{1},0;{\bf x}_{2},0)}\chi(x_{1}, x_{2})\nonumber\\
&&\qquad\qquad\qquad\qquad\qquad\qquad
\cdot\frac{1}{\sqrt{2}}e^{-is\sqrt{-\nabla_{1}^{2}+m^{2}}}\varphi^{\dagger}({\bf x}_{1}, t_{1}+s)e^{-is\sqrt{-\nabla_{2}^{2}+m^{2}}}\varphi^{\dagger}({\bf x}_{2}, t_{2}+s)|0\rangle\nonumber\\  
 &&=
\int dx_{1}dx_{2}e^{-is\big[\sqrt{-\nabla_{1}^{2}+m^{2}}+\sqrt{-\nabla_{2}^{2}+m^{2}}\big]}\Big[e^{-iW[({\bf x}_{1}-{\bf x}_{2})^{2}]\int_{0}^{s}dt\int d{\bf z}{\cal I}({\bf z},t;{\bf x}_{1},0;{\bf x}_{2},0)}\chi({\bf x}_{1}, t_{1}-s,{\bf x}_{2}, t_{2}-s )\Big]\nonumber\\
&&\qquad\qquad\qquad\qquad\qquad\qquad
\cdot\frac{1}{\sqrt{2}}\varphi^{\dagger}(x_{1})\varphi^{\dagger}(x_{2})|0\rangle.\nonumber\\                                          
\end{eqnarray}

To illustrate, as we have mentioned is evident in the interaction picture, this model has the peculiar property that, if the initial particle wave packets do not overlap over a region 
within the range of the potential $W$, there is no interaction.  How is this  evidenced in the Schr\"odinger picture?   

One sees in ({\ref{A3}) 
behavior similar to that associated with the non-relativistic  operator $W[({\bf X}_{1}-{\bf P}_{1}s/m-{\bf X}_{2}+{\bf P}_{2}s/m)^{2}]$.  ${\bf X}_{i}-{\bf P}_{i}s/m$ is essentially just  the initial position of a  packet.  Therefore, in the Schr\"odinger picture, the argument of the potential remains roughly constant, larger than the potential's range, even though the packets move toward each other, overlap and recede.  

Now we turn to consider how the energy-momentum of the particles is altered by the evolution ({\ref{A2}). In the limit $s\rightarrow\infty$, utilizing 
$\int_{0}^{s}dt\int d{\bf z}{\cal I}({\bf z},t;{\bf x}_{1},0;{\bf x}_{2},0))=C[{\bf x}_{1}-{\bf x}_{2}]^{4}$ ($C$ is a dimensionless constant whose value is unimportant for our discussion), 
\begin{equation}\label{A4}
\langle q_{1}, q_{2}|\psi,\infty\rangle_{I}\sim\int dx_{1}dx_{2}e^{-iC({\bf x}_{1}-{\bf x}_{2})^{4}W[({\bf x}_{1}-{\bf x}_{2})^{2}]}
\chi_{0}(x_{1}, x_{2})
e^{-iq_{1}\cdot{\bf x}_{1}}e^{-iq_{2}\cdot{\bf x}_{2}}
\end{equation}
\noindent plus a term with $q_{1}\leftrightarrow q_{2}$, which we shall hereafter omit writing.
Let the two particles have completely overlapping initial wave functions, $\chi_{0}(x_{1}, x_{2})=\chi_{0}(x_{1}) \chi_{0}(x_{2}) $, where $\chi_{0}(x_{j})$ is given by ({\ref{15}) :
\begin{equation}\label{A5}
\chi_{0}(x_{j})\sim e^{-\frac{t_{j}^{2}}{4\sigma'(2)}}\int d{\bf k}_{j}e^{-[{\bf k}_{j}-{\bf p}_{j}]^{2}\sigma^{2}}e^{i{\bf k}_{j}\cdot{\bf x}_{j}-i\omega({\bf k}_{j})t_{j}}.  
\end{equation}
\noindent Upon writing the potential term in ({\ref{A4}) as a Fourier transform, 
\begin{equation}\label{A6}
e^{-iC({\bf x}_{1}-{\bf x}_{2})^{4}W[({\bf x}_{1}-{\bf x}_{2})^{2}]}\equiv\int d{\bf k}e^{i{\bf k}\cdot({\bf x}_{1}-{\bf x}_{2})}g({\bf k}^{2}), 
\end{equation}
\noindent and inserting this and ({\ref{A5}) into ({\ref{A4}), we obtain
\begin{eqnarray}\label{A7}
\langle q_{1}, q_{2}|\psi,\infty\rangle_{I}&\sim&\int d{\bf k}d{\bf k}_{1}d{\bf k}_{2}\delta({\bf k}+{\bf k}_{1}-{\bf q}_{1})\delta(-{\bf k}+{\bf k}_{2}-{\bf q}_{2})g({\bf k}^{2})\nonumber\\
&&\qquad\qquad e^{-[{\bf k}_{1}-{\bf p}_{1}]^{2}\sigma^{2}}e^{-[{\bf k}_{2}-{\bf p}_{2}]^{2}\sigma^{2}}e^{-(q_{1}^{0}-\omega({\bf k}_{1}))^{2}\sigma'^{2}}e^{-(q_{2}^{0}-\omega({\bf k}_{2}))^{2}\sigma'^{2}}.
\end{eqnarray}

We see from the delta functions that momentum is conserved: for large enough $\sigma$, ${\bf q}_{1}+{\bf q}_{2}\approx {\bf p}_{1}+{\bf p}_{2}$, and ${\bf k}$ is the transferred momentum.  
Also, for large enough $\sigma'$, the particle energies are unchanged, $q_{j}^{0}\approx\omega({\bf k}_{j})\approx \omega({\bf p}_{j})$. (Our requirements that $\sigma'$ and $\sigma'^{-1}$ both be small do not conflict, since the former is a time,  e.g., $\sigma'\approx 7\times10^{-18}$sec, while the latter is an energy, e.g., $\hbar/\sigma'\approx 100$eV$\approx 10^{-7}m$.) 

However, the usual mass-shell relationships, which constrain the momentum transfer ${\bf k}$, do not exist here.  The outgoing masses of the particles depends upon the dynamics:
\begin{equation}\label{A8}
m_{j}^{2}\equiv q_{j}^{02}-{\bf q}_{j}^{2}\approx \omega^{2}({\bf p}_{j})-({\bf p}_{j}\pm{\bf k})^{2}=m^{2}\mp2{\bf p}_{j}\cdot{\bf k}-{\bf k}^{2}, 
\end{equation}
\noindent ${\bf k}$ can take on any value so, generally, there is a scattering to particles with a continuum distribution of masses. 

Still, one can arrange for the final mass spread to be 
of the order of accuracy to which the initial masses are known.  We note that if the only dimensional parameter characterizing the potential $W$ is its range $b$, then the Fourier transform $g({\bf k}^{2})$ 
will have the range $| {\bf k}|\approx 1/b$. So, for example, if we adopt the stringent constraint that the initial masses 
are known to $10^{-7}m\approx 100eV$, this will be approximately true also for the final masses 
if the initial momenta $p_{j}$ are such that the initial kinetic energy is $\lesssim 100$eV and also $\hbar^{2}/2mb^{2}\lesssim 100$eV, i.e., $ 5\times 10^{-11}$cm$\lesssim b$.

\section{Evaluation of integrals relevant to collapse calculation}

\subsection{$\ell=a$}

Here we evaluate the integral (\ref{31}):
\begin{equation}\label{B1}
I(T)\approx\Lambda D^{2}\int_{0}^{T}dx\int_{V} d{\bf z}_{1}\int_{V} d{\bf z}_{2}f[{\bf x}^{2}-t^{2}] f[({\bf x}- {\bf z})^{2}-t^{2}], 
\end{equation}
where $f(x^{2})=a^{-2}x^{2}\exp-(a^{-2}x^{2})\Theta(x^{2})=-\alpha\frac{\partial}{\partial \alpha}\exp-(\alpha x^{2})|_{\alpha=a^{-2}}\Theta(x^{2})$.  Because the integral is unchanged by replacement  ${\bf z}\rightarrow- {\bf z}$, it is symmetric in ${\bf x}\cdot{\bf z}=rz\cos\theta$, so its value is $2\times$ the integral over half the range of $\cos\theta$:
\begin{eqnarray}\label{B2}
I(T)&\approx&\Lambda D^{2}2\pi \int_{V} d{\bf z}_{1}\int_{V} d{\bf z}_{2}\alpha\frac{\partial}{\partial \alpha}\beta\frac{\partial}{\partial \beta}\int_{0}^{T}dt\int_{t}^{\infty}r^{2}dr
e^{-\alpha(r^{2}-t^{2})}e^{-\beta(r^{2}+z^{2}-t^{2})}2\int_{0}^{1} d\cos\theta e^{-\beta2 rz\cos\theta}\nonumber\\
&=&\Lambda D^{2}2\pi\int_{V} d{\bf z}_{1}\int_{V} d{\bf z}_{2}\frac{1}{z}\alpha\frac{\partial}{\partial \alpha}\beta\frac{\partial}{\partial \beta}e^{-\beta z^{2}}\int_{0}^{T}dt\int_{t}^{\infty}rdr
e^{-(\alpha+\beta)(r^{2}-t^{2})}\frac{1}{\beta} \Big[1- e^{-\beta rz}\Big].
\end{eqnarray}
\noindent We note that the constraint $r\geq t$ ensured that both exponents are positive over the positive range of integration of $\cos\theta$. 

We may drop $\exp-\beta rz$ compared to 1. The exponent $\beta rz\approx (z/a)(r/a)$. The first factor $ z/a\approx 1$ for most of the range of integration over $z$  on account  of the factor $\exp-\beta z^{2}$.  However,  since  $a\approx 3\times10^{-16}$s, then $t/a>>1$ for most of the range of $t$, so  $r/a>t/a>>1$.

The integrals over $r$ and $t$ are then readily performed, followed by the derivatives with respect to $\alpha, \beta$, and then replacement of these variables by $a^{-2}$:
\begin{eqnarray}\label{B3}
I&\approx&\pi T\Lambda D^{2}\int_{V} d{\bf z}_{1}\int_{V} d{\bf z}_{2}\frac{1}{z}\alpha\frac{\partial}{\partial \alpha}\beta\frac{\partial}{\partial \beta}\frac{1}{(\alpha+\beta)\beta}e^{-\beta z^{2}}\nonumber\\
&=&\Lambda T\frac{\pi a^{2}}{4}D^{2}\int_{V} d{\bf z}_{1}\int_{V} d{\bf z}_{2}\frac{1}{z}[z^{2}+2a^{2}]e^{-a^{-2}z^{2}}.
\end{eqnarray}
The remaining integrals can be performed by replacing the integration variable ${\bf z}_{2}$ by ${\bf z}={\bf z}_{2}-{\bf z}_{1}$: one can extend the integration range of $z$ to infinity because the 
exponent effectively limits the range to a small volume of scale $a$ and we have assumed that $V>>a^{3}$:
\begin{eqnarray}\label{B4}
I(T)&\approx&\Lambda T\frac{\pi a^{2}}{4}D^{2}\int_{V} d{\bf z}_{1}4\pi\int_{0}^{\infty}z^{2}dz\frac{1}{z}[z^{2}+2a^{2}]e^{-a^{-2}z^{2}}\nonumber\\
&=&\Lambda T\frac{\pi a^{2}}{4}D^{2}\int_{V} d{\bf z}_{1}6\pi a^{4}=\frac{3\pi^{2}}{2}\Lambda Ta^{3}[DV][Da^{3}].
\end{eqnarray}

\subsection{$\ell=$ size of universe}

In Eq.(\ref{29}) , we may set $f[({\bf x}- {\bf z})^{2}-t^{2}]\approx f[{\bf x}^{2}-t^{2}]$  since $z<<|{\bf x}|\simeq \ell$:
\begin{eqnarray}\label{B5}
I(T)&\approx&\Lambda D^{2}\int_{0}^{T}dx\int_{V} d{\bf z}_{1}\int_{V} d{\bf z}_{2}f[{\bf x}^{2}-t^{2}]\Big[ f[{\bf x}^{2}-t^{2}] 
  - f[({\bf x}-{\bf d})^{2}-t^{2}]\Big].\nonumber\\
  &=&\Lambda D^{2}V^{2}\int_{0}^{T}dt\int d{\bf x}\frac{x^{2}}{\ell^{4}}e^{-\frac{2x^{2}}{\ell^{2}}}\Big[x^{2}-(x^{2}-2{\bf x}\cdot{\bf d}+d^{2})
  e^{-\frac{-2{\bf x}\cdot{\bf d}+d^{2}}{\ell^{2}}}\Big].
\end{eqnarray}
Since $T<<\ell/c$, then $x^{2}\approx {\bf x}^{2}\equiv r^{2}$ and the time integral gives $T$. The exponent in the second term in the square bracket $<<1$ can be expanded, and the integral over ${\bf x}\cdot {\bf d}$ vanishes. The resulting expression, neglecting $d^{2}/\ell^{2}$ compared to 1 is
\begin{eqnarray}\label{B6}
I(T)&\approx&\Lambda n^{2}T2\pi\int d{\bf x}\frac{r^{2}}{\ell^{4}}e^{-2\frac{r^{2}}{\ell^{2}}}\Big[4\frac{({\bf x}\cdot{\bf d})^{2}}{\ell^{2}}-d^{2}\Big]\nonumber\\
&\approx&
\Lambda n^{2}Td^{2}2\pi\int_{0}^{\infty} dr\frac{r^{4}}{\ell^{4}}e^{-2\frac{r^{2}}{\ell^{2}}}\Big[\frac{8r^{2}}{3\ell^{2}}-1\Big]=\frac{\pi}{2}\Lambda n^{2}Td^{2}\ell.
\end{eqnarray}

\end{document}